\documentclass[prx,twocolumn,nofootinbib,longbibliography,preprintnumbers,amsmath,amssymb,10pt]{revtex4}

\usepackage{bm,graphicx,multirow,subfigure,cases}

\DeclareMathOperator{\sgn}{sgn}

\newcommand{\numberfield}[1]{\ensuremath{\mathbb{#1}}} 

\newcommand{\mat}[1]{\ensuremath{\mathbf{#1}}} 

\newcommand{\Bpi}{{\bm{\pi}}} 

\newcommand{\Btau}{{\bm{\tau}}} 

\newcommand{\group}[1]{\ensuremath{#1}} 

\newcommand{\cunitdisc}{\ensuremath{\overline{\mathbb{D}}}} 
\newcommand{\chunitdisc}{\ensuremath{\overline{\mathbb{D}}}_{1/2}} 

\newcommand{\set}[1] {\ensuremath{\mathcal{#1}}} 

\newcommand{\mment}[1]{\ensuremath{\mathbf{#1}}} 

\newcommand{\cmment}[1]{\ensuremath{\widetilde{\mathbf{#1}}}} 

\newcommand{\sys} {\ensuremath{\mathcal{S}}} 

\newcommand{\seq}[2]{\ensuremath{[#1,#2]}}   

\newcommand{\lseq}[3]{\ensuremath{[#1,#2,#3]}}   

\newcommand{\llseq}[4]{\ensuremath{[#1,#2,#3,#4]}}   

\newcommand{\cout}[2]{\ensuremath{(#1\colon#2)}}
\newcommand{\coutd}[3]{\ensuremath{(#1\colon#2\colon\dots\colon#3)}}
\newcommand{\coutdd}[4]{\ensuremath{(#1\colon#2\colon\dots\colon#3\colon\dots\colon#4)}}

\newcommand{\sseq}[1]{\ensuremath{#1}}    

\newcommand{\bubble}[2]{\ensuremath(#1,#2)}   

\newcommand{\pll}{\lor}  

\newcommand{\ser}{\ensuremath{\mathop{\bm{\cdot}}}}  

\newcommand{\amp}{z}  

\newcommand{\prob}[1]{\ensuremath{P(\sseq{#1})}}   

\newcommand{\revsseq}[1]{\ensuremath\overleftarrow{#1}}    

\newcommand{\nsys}{\ensuremath{N}} 

\newcommand{\Gmat}[4]{\ensuremath G(#1,#2, #3, #4) } 
\newcommand{\GMat}[4]{\ensuremath G\begin{pmatrix} #1, #2, \quad \\ \quad #3, #4 \end{pmatrix}}

\begin{document}


\title{Informational Approach to the Symmetrization Postulate}
\author{Philip Goyal}	
    \email{pgoyal@albany.edu}
    \affiliation{University at Albany~(SUNY), NY, USA}
\date{\today}
\begin{abstract}
 
A remarkable feature of quantum theory is that particles with identical intrinsic properties must be treated as indistinguishable if the theory is to give valid predictions.  In the quantum formalism, indistinguishability is expressed via the symmetrization postulate~\cite{Dirac26, Heisenberg26}, which restricts a system of identical particles to the set of symmetric states (`bosons') or the set of antisymmetric states~(`fermions').  However, the physical basis and range of validity of the symmetrization postulate has not been established. A well-known topological derivation of the postulate implies that its validity depends on the dimensionality of the space in which the particles move~\cite{LaidlawDeWitt71,LeinaasMyrheim77}.   However, this derivation relies on the labeling of indistinguishable particles, a notion that cannot be justified on an informational basis.   Here we show that the symmetrization postulate can be derived by strictly adhering to the informational requirement that particles which cannot be experimentally distinguished from one another are not labeled.  Our key novel postulate is the \emph{operational indistinguishability postulate,} which posits that the amplitude of a process involving several indistinguishable particles is determined by the amplitudes of all possible transitions of these particles when treated as distinguishable. The symmetrization postulate follows by requiring consistency with the rest of the quantum formalism.    The derivation implies that the symmetrization postulate admits no natural variants.  In particular, the possibility that identical particles generically exhibit anyonic behavior in two dimensions is excluded.

\end{abstract}

\maketitle

The mathematical formalism of quantum theory has numerous features whose physical origin is obscure.   Although these features have thus far proven to be empirically adequate, a proper understanding of their physical origin in vital in order that their range of validity be established.  In recent years, there has been growing appreciation that an informational viewpoint provides a powerful vantage point for constructing such an understanding~\cite{Wheeler89, Zeilinger99, Fuchs02,Grinbaum-reconstruction}, and substantial progress has now been made, both in deriving much of the quantum formalism from informational principles~\cite{Hardy01a, Clifton-Bub-Halvorson03, DAriano-operational-axioms, Goyal-QT, Reginatto-Schroedinger-derivation, GKS-PRA, Goyal2014a, Chiribella2011}, and in identifying informational principles that account for some of the nonclassical features of quantum physics~\cite{Barrett2006,information-causality}.

However, the quantum treatment of identical particles, one of the most important features of quantum theory insofar as its application is concerned, has hitherto not been analyzed from an informational standpoint.  Our purpose here is to provide such an analysis.

In classical physics, identical particles can be treated as distinguishable since one can, in principle, follow the precise trajectory of each particle over time.  However,  in order to yield predictions in accordance with experiment, it is essential that quantum theory treat identical particles as \emph{indistinguishable}.  Formally, the indistinguishability of the particles in a quantum system is expressed via the \emph{symmetrization postulate}~\cite{Dirac26, Heisenberg26}, which restricts the system to symmetric states or to antisymmetric states.  This postulate is responsible for the classification of all identical particles in nature into bosons and fermions, and is required for the understanding of a vast range of physical phenomena.

Shortly after its inception, it was recognized that the physical basis, and thus the sphere of validity, of the symmetrization postulate is uncertain~\cite{Dirac30,PauliWellenmechanik33}, and there have since been numerous efforts to elucidate its physical basis and range of validity~\cite{MG64, Girardeau65, Steinmann66, LaidlawDeWitt71, LeinaasMyrheim77, Tikochinsky-identical, Tikochinsky-Shalitin, Bacciagaluppi2003, Durr2006, Peshkin2003, NeoriGoyal2012}.  The most widely known physical explanation of the postulate is the so-called topological approach~\cite{LaidlawDeWitt71, LeinaasMyrheim77}.  Remarkably, the topological approach implies that the symmetrization postulate is not unconditionally valid, its validity depending upon the spatial dimension in which identical particles move.  In three dimensions, only fermions and bosons are possible, in accord with the symmetrization postulate; but, in two spatial dimensions, the topological approach predicts that identical particles can also exhibit so-called anyonic behavior.  This prediction has been widely influential in fields as diverse as condensed matter physics~\cite{Stern2007} and topological quantum computing~\cite{Anyons-Topological-Quantum-Computation}, and is commonly cited as a rationale for experimental investigation of anyonic behavior.

However, when viewed from an informational point of view, the topological approach suffers from a crucial flaw.  A vital feature of the informational viewpoint is its insistence that theoretical distinctions be underpinned by experimentally measurable data.  Such a stipulation helps to guard against implicit assumptions.  However, the topological approach assigns labels to particles that are supposed to be indistinguishable.  But, since the particles are indistinguishable, such labeling cannot have any basis in an experimental fact, and is therefore inadmissible from an informational point of view.  As we show elsewhere~\cite{Goyal-anyons}, this labeling indeed conceals a nontrivial implicit assumption that has not hitherto been noticed.

Here we show that the symmetrization postulate can be derived in full generality within a fully informational approach:~if two particles are indistinguishable on the basis of information obtainable through measurements on them, we refrain from labeling the particles.  Our key postulate is the  \emph{operational indistinguishability postulate}, which establishes a relation between the theoretical description of two different experiments, positing that the amplitude of a process involving several indistinguishable subsystems (hereafter referred to as~\emph{particles}) is determined by the amplitudes of all possible transitions of these particles when treated as distinguishable.   Our informational approach allows us to clearly grasp the essential meaning of the symmetrization postulate, and to steer clear of conceptual pitfalls that commonly afflict treatments of this subject~\cite{Mirman73, Muynck75, Brown99}. 

The derivation harnesses the framework and methodology recently used to derive Feynman's formalism of quantum theory~\cite{GKS-PRA}, in two key ways.  First, we make use of the insight that Feynman's rules are best regarded as establishing a formal relationship between different experiments~\cite{GKS-PRA, Goyal-Knuth2011}, an insight that lies at the basis of the indistinguishability postulate.  Second, we employ the simple yet powerful~\cite{GKS-PRA, Goyal-Knuth2011} requirement of consistency:~if it is possible to compute an amplitude in two different ways, they must agree.

\section*{Two Indistinguishable Particles}
\label{sec:two-identical}

\begin{figure}
\begin{center}
\includegraphics[width=3.25in]{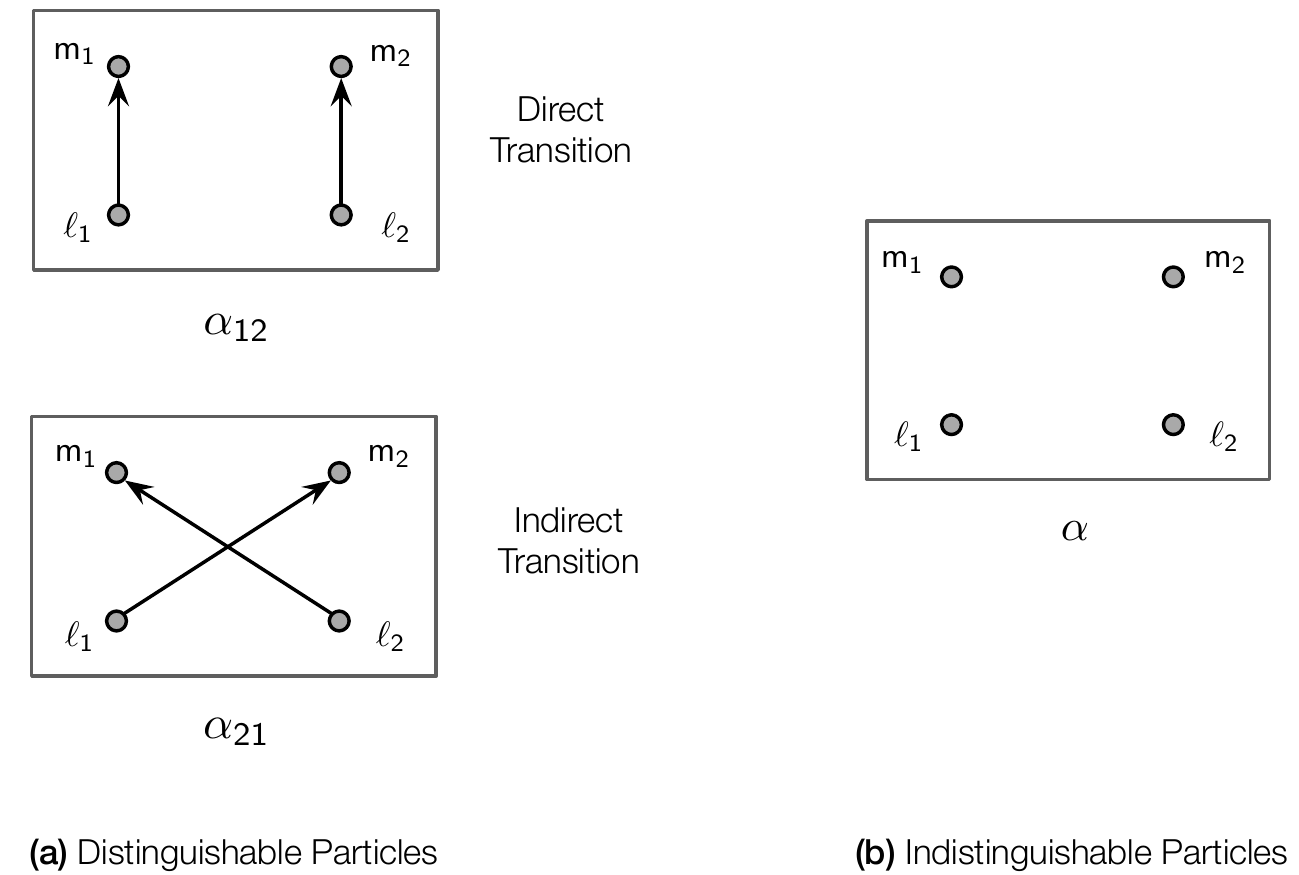}
\caption{\label{fig:distinguishable-and-indistinguishable}\emph{Measurements on two particles, distinguishable and indistinguishable.}  Two particles are each subject to measurements at times,~$t_1$ and~$t_2$, yielding outcomes~$\ell_1, \ell_2$ at~$t_1$ and~$m_1, m_2$ at~$t_2$.  \emph{Left:} If the particles are distinguishable, one can experimentally determine which transition actually occurs---the `direct' one~(\emph{top}), or the `indirect' one~(\emph{bottom}).  \emph{Right:} If the particles are indistinguishable, one cannot say what happened in the intermediate time on the basis of the information experimentally obtained.}
\end{center}
\end{figure}

Consider a quantum system of two particles, each subject to measurements~(say, of position) at successive times~$t_1$ and~$t_2$.  Suppose that the two measurements performed at~$t_1$ yield outcomes~$\ell_1$ and~$\ell_2$, and that the measurements at~$t_2$ yield outcomes~$m_1$ and~$m_2$.  If it is possible to distinguish between the two particles, then there is some measurement~(for example, of mass or charge) that we can, in principle, perform on each particle immediately before~$t_1$ and after~$t_2$ that allows us to conclude, for instance, that the particle responsible for outcome~$\ell_1$ is the same as that responsible for~$m_1$~(see Fig.~\ref{fig:distinguishable-and-indistinguishable}). 
If, however, the particles cannot be distinguished from one another, we cannot draw such a conclusion.  Even the conclusion that ``the particle that generated outcome~$\ell_1$ was also responsible for outcome~$m_1$ or $m_2$, but we do not know which'' is  invalid since we cannot verify this statement on account of particle indistinguishability.  To nonetheless draw such a conclusion would be tantamount to assuming the separate and continued existence of the particles between observations, which assumption our experience with quantum phenomena should warn us against.  In particular, in the case of electron diffraction at a double-slit~\cite{FeynmanHibbs65,Goyal-Knuth2011}, in the absence of which-way detectors at each slit, we cannot infer from the experimental data that the electron passed through one slit or the other; if we nonetheless insist on drawing this inference, our predictions are incorrect---we are unable to account for the observed diffraction pattern.

\begin{figure}
\begin{center}
\subfigure[]
				{
				\includegraphics[width=3.5in]{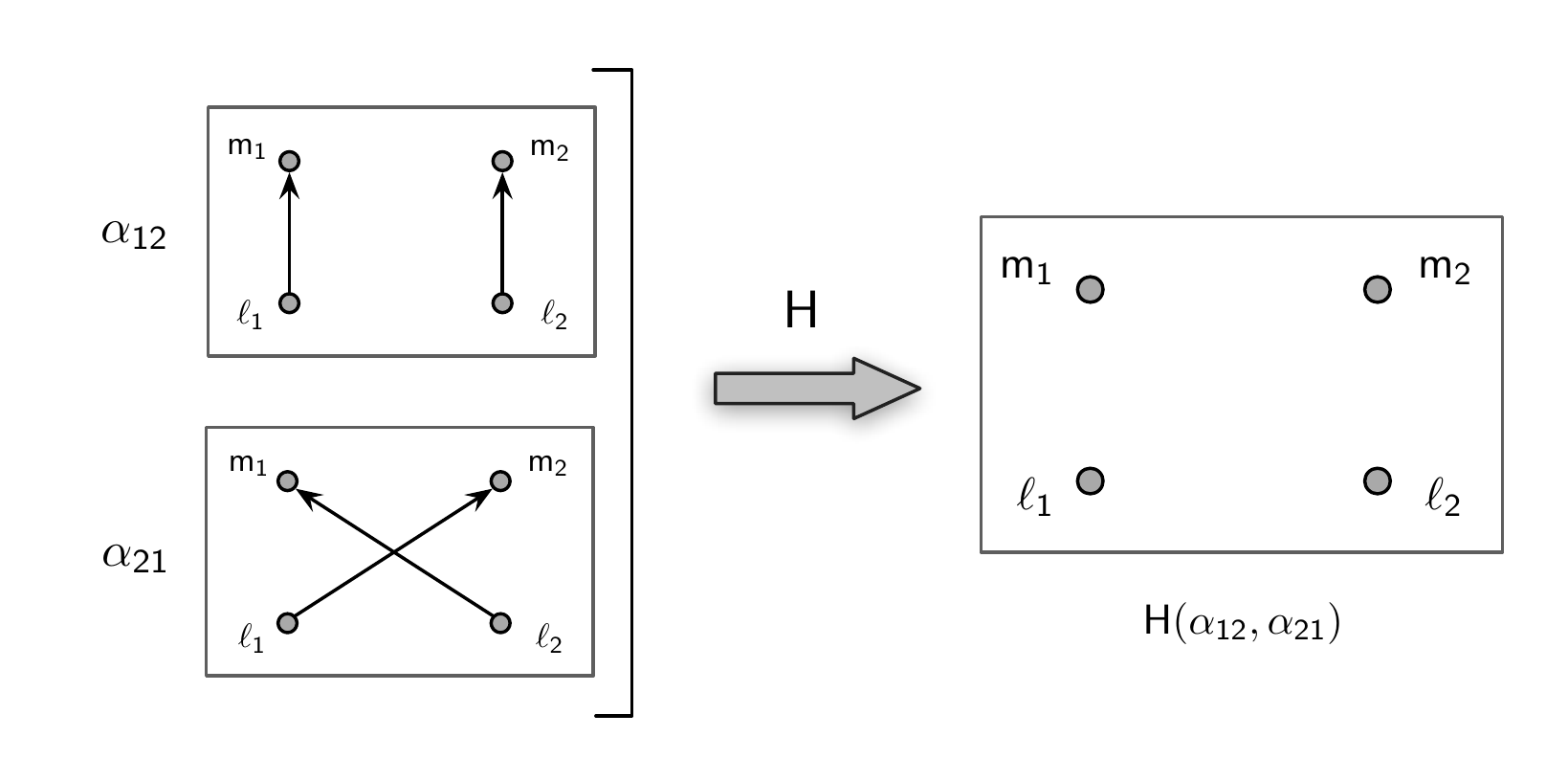}
				\label{fig:two-particles-one-stage}
				}
\subfigure[]
				{
				\includegraphics[width=3.25in]{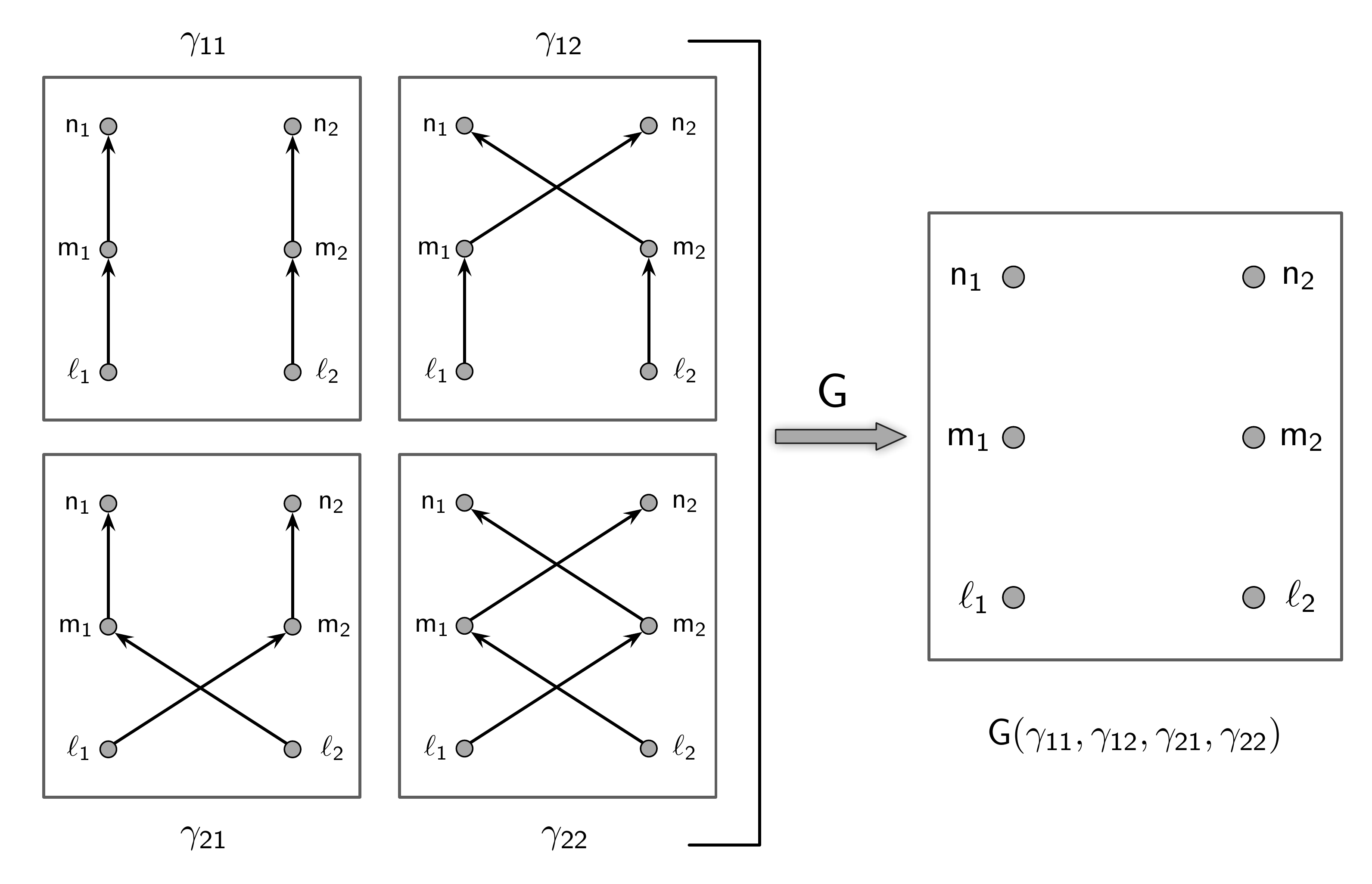}
				\label{fig:two-particles-two-stages}
				}
\caption{\emph{Amplitude for one- and two-stage experiments on two indistinguishable particles.}  Two indistinguishable particles are each subject to a measurement at successive times.  (a)~Measurements at time~$t_1$ and~$t_2$ yield outcomes~$\ell_1, \ell_2$ and~$m_1, m_2$, respectively.   The figures on the left show the transitions of two \emph{distinguishable} particles compatible with these outcomes:~the `direct' transition of amplitude~$\alpha_{12}$, and the `indirect' transition of amplitude~$\alpha_{21}$.  We postulate that the amplitude of the indistinguishable-particle process is~$H(\alpha_{12}, \alpha_{21})$, where~$H$ is a continuous function to be determined.  (b)~Measurements are performed at three successive times, yielding the indicated outcomes.  On the left are shown the four possible transitions of two distinguishable particles, with respective amplitudes~$\gamma_{11}, \gamma_{12}, \gamma_{21}, \gamma_{22}$, compatible with the observed outcomes.  We postulate that the amplitude of the indistinguishable-particle process is~$\Gmat{\gamma_{11}}{\gamma_{12}}{\gamma_{21}}{\gamma_{22}}$.}
\end{center}
\end{figure}

As we have no way of `looking inside' the indistinguishable-particle process, we must treat it as a black box.  As a result, we cannot directly compute the amplitude,~$\alpha$, of the indistinguishable-particle process.  Again, electron diffraction is analogous:~the transition amplitude of an electron from source to screen in the absence of which-way detectors cannot be computed directly.  Instead, it is obtained~(via Feynman's amplitude sum rule) from the sum of the amplitudes of the two possible transitions through the slits with the detectors present, which amplitudes \emph{can} be calculated~\cite{Goyal-Knuth2011}.  In the present case of two particles, then, we seek to relate the amplitude of the indistinguishable-particle transition,~$\alpha$, with the amplitudes,~$\alpha_{12}$ and~$\alpha_{21}$, of the `direct'~(one particle moves from~$\ell_1$ to~$m_1$, the other from~$\ell_2$ to~$m_2$) and `indirect'~(one particle moves from~$\ell_1$ to~$m_2$, the other from~$\ell_2$ to~$m_1$) transitions of the two distinguishable particles, where each of these particles agree in all their dynamically-relevant properties with the indistinguishable particles.   Now, we cannot relate~$\alpha$ to~$\alpha_{12}$ and~$\alpha_{21}$ via the sum rule since the sum rule only applies when different experiments are applied to the \emph{same} system, whereas~$\alpha$ refers to indistinguishable particles, while~$\alpha_{12}$ and~$\alpha_{21}$ refer to distinguishable particles.  So, in order to derive a connection between~$\alpha$ and~$\alpha_{12}$ and~$\alpha_{21}$, we must first assume that, as in the case of electron diffraction, such a relation exists.   Accordingly, we postulate that
\begin{equation} \label{eqn:H-def}
\alpha = H(\alpha_{12}, \alpha_{21}),
\end{equation}
where~$\alpha$ lies in the closed unit disc,~$\cunitdisc$, in the complex plane, and~$H$ is a continuous function, with some domain~$S = H^{-1}(\cunitdisc)  \subseteq \cunitdisc \times \cunitdisc$, to be determined~(see Fig.~\ref{fig:two-particles-one-stage}).  Similarly, when two particles are subject to three successive measurements, we postulate that the indistinguishable-particle amplitude,~$\gamma \in \cunitdisc$, is given by
\begin{equation} 
\label{eqn:G-def}
\gamma = \Gmat{\gamma_{11}}{\gamma_{12}}{\gamma_{21}}{\gamma_{22}},
\end{equation}
where~$\gamma_{11}, \gamma_{12}, \gamma_{21},$ and~$\gamma_{22}$ are the amplitudes of the corresponding distinguishable-particle transitions~(see Fig.~\ref{fig:two-particles-two-stages}), and~$G$ is a function, with some domain~$G^{-1}(\cunitdisc)$, to be determined.  
Both of these postulates are special cases of the operational indistinguishability postulate stated above.

\begin{figure}
\begin{center}
\includegraphics[width=3.5in]{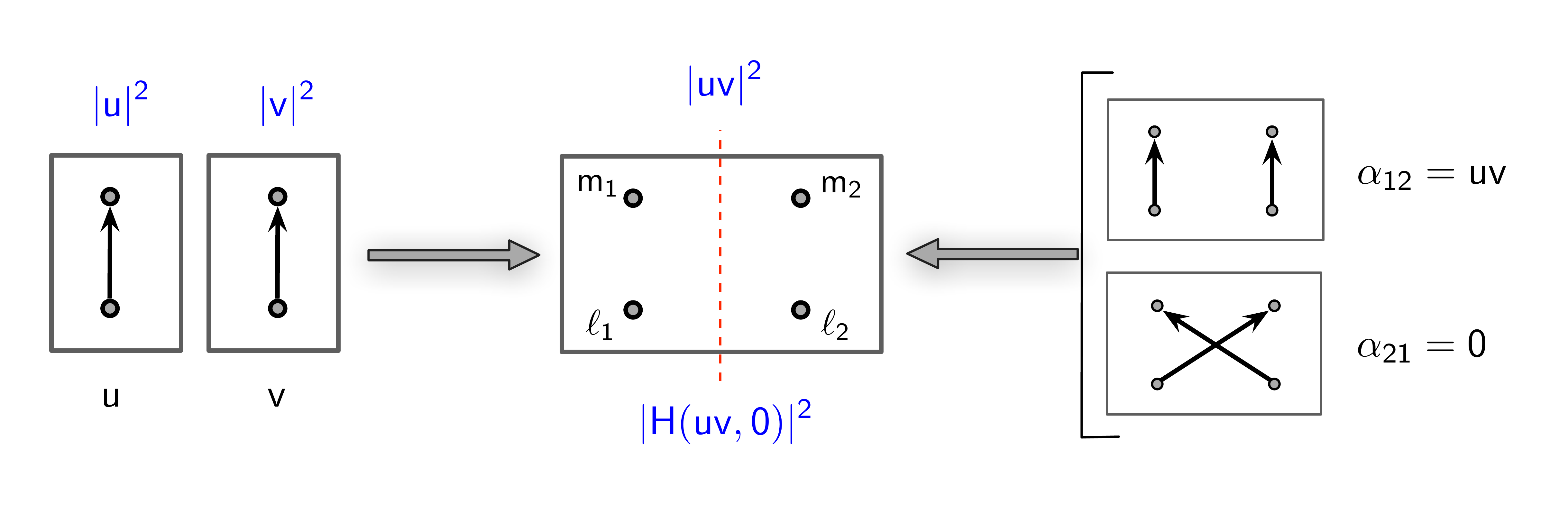}
\caption{\label{fig:isolation-two-particles}  \emph{Isolation condition.}  In the center, two indistinguishable particles are each subject to measurements at times~$t_1$ and $t_2$, yielding the indicated outcomes.  Each particle is bound to its own subexperiment, which are isolated from one another.  Accordingly, the transition probability can be computed in two different ways.  As on the left, one can compute the transition probabilities,~$|u|^2$ and~$|v|^2$, for each of the particles separately, and multiply these to obtain~$|uv|^2$.  Alternatively, as on the right, one can use the~$H$ function to compute the amplitude~$H(\alpha_{12}, \alpha_{21}) = H(uv, 0)$, which yields transition probability~$|H(uv, 0)|^2$.  Consistency requires that~$|H(uv, 0)|^2 = |uv|^2$.}
\end{center}
\end{figure}

Additionally, consider the special case that each particle is known to be bound to a separate experiment, and that the particles do not interact with one another~(see Fig.~\ref{fig:isolation-two-particles}).
That is, each particle is confined to its own \emph{subexperiment,} and the two subexperiments are physically isolated from one other.  In this case, the particle originating at~$\ell_1$ will later be detected at, say,~$m_1$, while the particle originating at~$\ell_2$ will be detected at~$m_2$.    As an example, let one of the particles be an electron that is bound to a proton in a lab on the earth, while the other be, similarly, an electron bound to a proton in a lab on the moon.  In such a case, each subexperiment is statistically independent of the other.  That is, the outcome probabilities observed in one subexperiment are indepen1dent of the outcome probabilities of the other subexperiment, or even of whether or not the other subexperiment is performed.   

In this case of isolated subexperiments, we assume, as is conventional, that we can compute the transition probabilities of each subexperiment without regard for the other.  We refer to this as the \emph{isolation condition}.  In particular, if the amplitude of the transition from~$\ell_1$ to~$m_1$ is~$u$, then the transition probability is simply~$|u|^2$. Likewise, if the amplitude of the transition from~$\ell_2$ to~$m_2$ is~$v$, then the transition probability is~$|v|^2$.  Since the subexperiments are statistically independent, the overall transition probability~(initial detection at~$\ell_1$ and~$\ell_2$, final detection at~$m_1$ and~$m_2$) is~$|u|^2 \cdot |v|^2 = |uv|^2$.

Now, we can also compute this overall transition probability using~$H$.  In this special case,~$\alpha_{21} = 0$ but~$\alpha_{12} = uv$.   Hence, the overall transition amplitude is simply~$H(uv, 0)$, which yields an overall transition probability of~$\left| H(uv, 0) \right|^2$.  Consistency requires that this agree with the overall transition probability computed above, so that, for all~$z \in \cunitdisc$,
\begin{equation} \label{eqn:normalization1}
\left| H(z, 0) \right| = |z|.
\end{equation}
Similarly, if the measurements~$\mment{L}_1$ and~$\mment{M}_2$ comprise one subexperiment, and measurements~$\mment{L}_2$ and~$\mment{M}_1$ another, and these subexperiments are isolated from one another, then, for all~$z \in \cunitdisc$,
\begin{equation} \label{eqn:normalization2}
\left| H(0,z) \right| = |z|.
\end{equation}

We shall now show that these assumptions determine the precise form of symmetrization postulate for two particles.  The general case of~$N$ particles, together with additional detail and background terminology, can be found in Appendix~\ref{sec:many-identical}.  

To determine~$G$ and~$H$, we repeatedly use the idea that, if it is possible to compute an amplitude in two different ways using Feynman's rules for single systems~\cite{Feynman48, GKS-PRA}~(see also Appendix~\ref{sec:Feynmans-rules}) and the assumptions given above, consistency requires that they be equal.  Each such call for consistency yields a functional equation.  By considering a few special cases, we are able to formulate a set of functional equations that determine~$G$ and~$H$.  

\begin{figure}
\begin{center}
\includegraphics[width=3.5in]{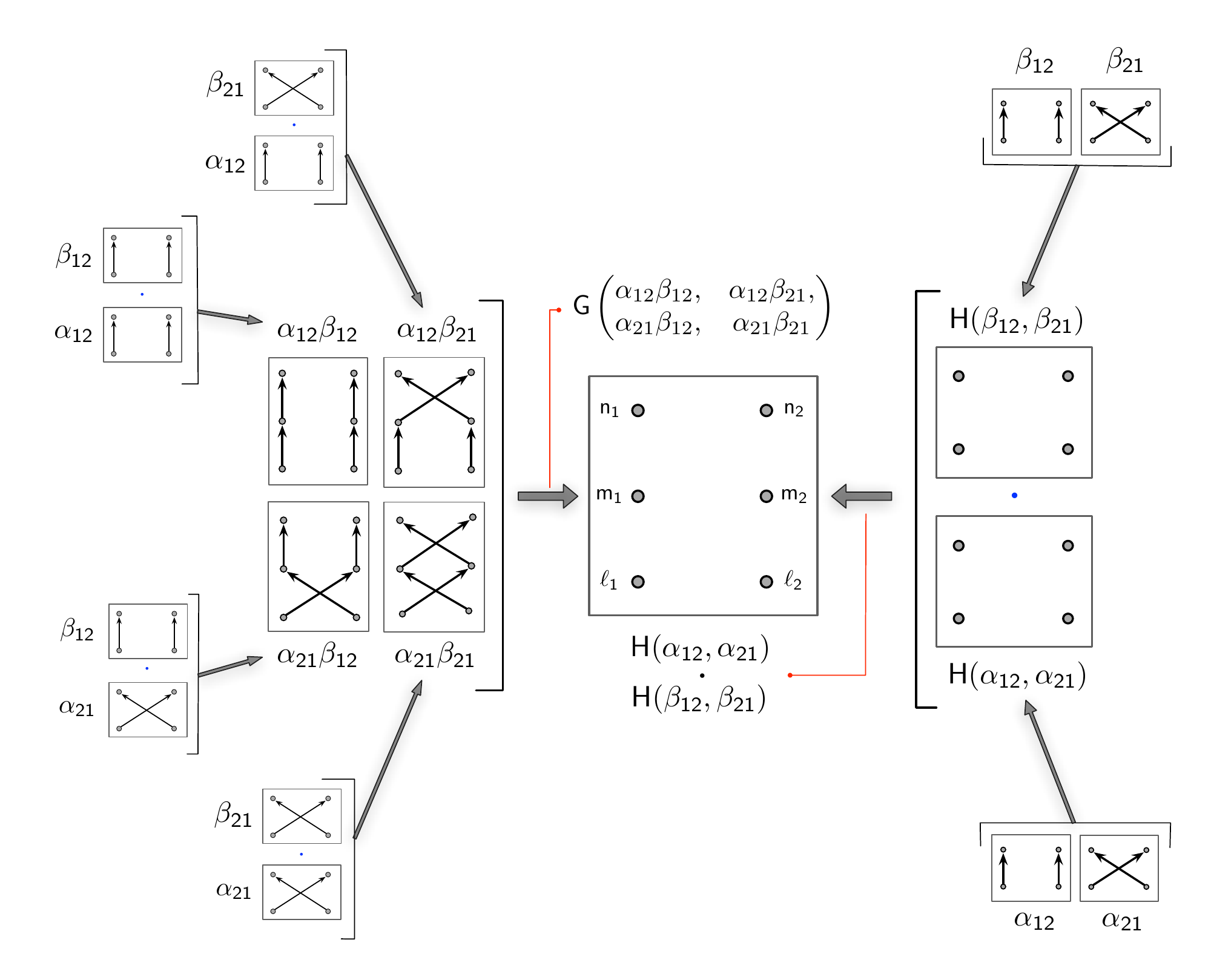}
\caption{\label{fig:two-particles-G-product}  \emph{Origin of the $G$-product equation.}  In the center, two indistinguishable particles are each subject to measurements at times~$t_1$, $t_2$ and~$t_3$, yielding the indicated outcomes. The amplitude for this process can be computed in two different ways.  On the left are shown the four possible transitions of two distinguishable particles compatible with the observed sequence.  These transitions can each be expressed as series~$(\ser)$ combinations of the indicated transitions, yielding amplitudes~$\alpha_{12}\beta_{12}, \alpha_{12}\beta_{21}, \alpha_{21}\beta_{12}, \alpha_{21}\beta_{21}$.  Hence, from Eq.~\eqref{eqn:G-def}, the indistinguishable-particle process amplitude is~$\Gmat{\alpha_{12}\beta_{12}}{\alpha_{12}\beta_{21}}{\alpha_{21}\beta_{12}}{\alpha_{21}\beta_{21}}$.  On the right, the indistinguishable-particle process is first decomposed into two smaller indistinguishable-particle processes in series~$(\ser)$, with respective amplitudes,~$H(\alpha_{12}, \alpha_{21})$ and~$H(\beta_{12}, \beta_{21})$, yielding an overall indistinguishable-particle amplitude~$H(\alpha_{12}, \alpha_{21}) \, H(\beta_{12}, \beta_{21})$.}
\end{center}
\end{figure}

First, consider a two-stage experiment in which the intermediate outcomes~$m_1$ and~$m_2$ are both atomic~\cite{GKS-PRA}~(see Appendix~\ref{sec:many-identical} for terminology).  We can derive the indistinguishable-particle amplitude,~$\gamma$, in two different ways~(see Fig.~\ref{fig:two-particles-G-product}). 
Let~$\alpha_{12}, \alpha_{21}$, respectively, be the amplitudes of the direct and indirect transitions in the first stage, and~$\beta_{12}, \beta_{21}$ the corresponding amplitudes in the second stage.  Using the amplitude product rule, we can directly compute the values~$\gamma_{11} = \alpha_{12}\beta_{12}$, $\gamma_{12} =\alpha_{12}\beta_{21}$, $\gamma_{21} =\alpha_{21}\beta_{12}$, and~$\gamma_{22} =\alpha_{21}\beta_{21}$. Then, using the~$G$-function defined above,
\begin{equation*}
\gamma = \Gmat{\alpha_{12} \beta_{12}}{\alpha_{12}\beta_{21}}{\alpha_{21}\beta_{12}} {\alpha_{21}\beta_{21}}.
\end{equation*}
Alternatively, we can directly apply the amplitude product rule to the system of identical particles to obtain
\begin{equation*}
\gamma = H (\alpha_{12}, \alpha_{21}) \, H(\beta_{12}, \beta_{21}).
\end{equation*}
Equating these two expressions, we obtain a functional equation, the $G$-product equation,
\begin{multline} \label{eqn:G-product}
\Gmat{\alpha_{12} \beta_{12}}{\alpha_{12}\beta_{21}}{\alpha_{21}\beta_{12}} {\alpha_{21}\beta_{21}}  = \\ H (\alpha_{12}, \alpha_{21}) \, H(\beta_{12}, \beta_{21}),
\end{multline}
valid for all~$(\alpha_{12}, \alpha_{21}), (\beta_{12}, \beta_{21}) \in S$.

\begin{figure*}
\begin{center}
\includegraphics[width=7in]{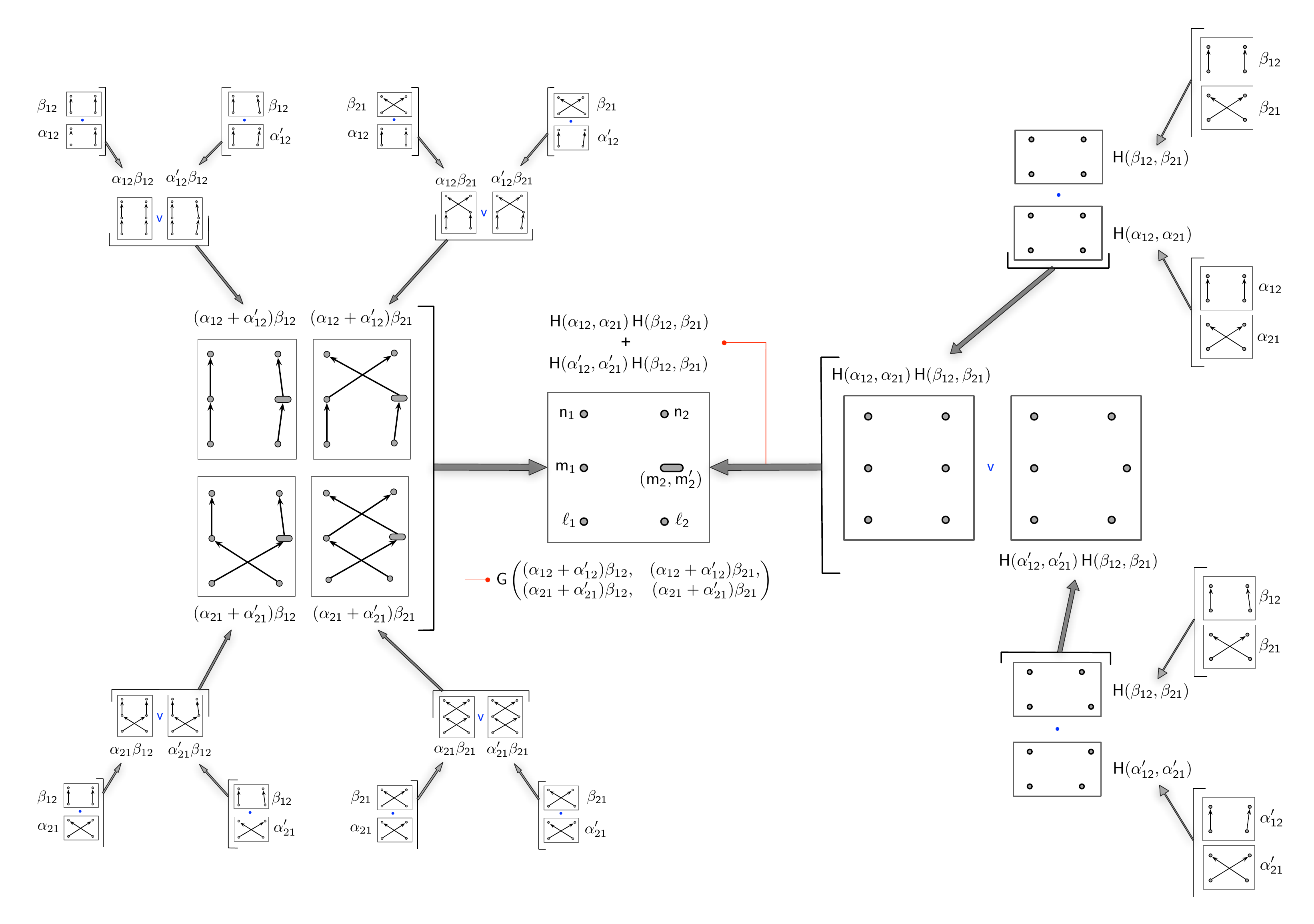}
\caption{\label{fig:two-particles-coarse-graining}  In the center, a system of two indistinguishable particles is subject to measurements at three successive times.  One of the outcomes of the measurements at~$t_2$ is a coarse-graining of atomic outcomes~$m_2$ and~$m_2'$, denoted~$\bubble{m_2}{m_2'}$.   The amplitude of this indistinguishable-particle process can be computed in two different ways: (i) on the right, the process is decomposed into two processes combined in parallel~$(\pll)$, and each of these in turn is expressed as two processes combined in series~$(\ser)$.  The amplitudes of each of the four component processes can then be determined using~$H$ in terms of the amplitudes of distinguishable-particle transitions, yielding~$[H(\alpha_{12}, \alpha_{21}) + H(\alpha_{12}', \alpha_{21}')]\,H(\beta_{12}, \beta_{21})$; (ii)~on the left are shown the four compatible transitions of two distinguishable particles.   The amplitudes of these transitions, ~$\gamma_{11} = (\alpha_{12} + \alpha_{12}')\beta_{12}, \gamma_{12} = (\alpha_{12} + \alpha_{12}')\beta_{21}, \gamma_{21} = (\alpha_{21} + \alpha_{21}')\beta_{12}$, and~$\gamma_{22} = (\alpha_{21} + \alpha_{21}')\beta_{21}$, are computed using Feynman's sum and product rules as indicated.   The amplitude of the indistinguishable-particle sequence is then obtained using the $G$-function to be~$\Gmat{\gamma_{11}}{\gamma_{12}}{\gamma_{21}}{\gamma_{22}}$.}
\end{center}
\end{figure*}

Next, consider a two-stage experiment in which the outcome of one of the measurements performed at~$t_2$ is a coarse-graining of atomic outcomes~$m_2$ and~$m_2'$, which outcome we denote~$\bubble{m_2}{m_2'}$~(see Fig.~\ref{fig:two-particles-coarse-graining}).  
Here,~$\alpha_{12}, \alpha_{21}$ are respectively the first-stage amplitudes of the direct and indirect transitions compatible with the indistinguishable-particle process through~$m_2$; $\alpha_{12}', \alpha_{21}'$ are similarly the amplitudes corresponding to the process through~$m_2'$; and~$\beta_{12}, \beta_{21}$ are the second-stage amplitudes of the direct and indirect transitions compatible with the indistinguishable-particle process through both~$m_2$ and~$m_2'$.  

The indistinguishable-particle amplitude,~$\tilde{\gamma}$, in this case can be computed in two different ways.  First, using the amplitude sum and product rules, we can write
\begin{equation*} \label{eqn:coarse-grain-amp1}
\tilde{\gamma} = H(\alpha_{12}, \alpha_{21}) \, H(\beta_{12}, \beta_{21}) + H(\alpha'_{12}, \alpha'_{21}) \, H(\beta_{12}, \beta_{21}),
\end{equation*}
Alternatively, one can compute~$\tilde{\gamma}$ directly from Eq.~\eqref{eqn:G-def}:
\begin{equation*} \label{eqn:coarse-grain-amp2}
\tilde{\gamma} = \GMat{(\alpha_{12}+ \alpha_{12}')\beta_{12}}{(\alpha_{12}+ \alpha_{12}')\beta_{21}}{(\alpha_{21} + \alpha_{21}')\beta_{12}}{(\alpha_{21} + \alpha_{21}')\beta_{21}},
\end{equation*}
where we have used the amplitude sum rule to compute the amplitude of the four transitions compatible with the indistinguishable-particle process.  Hence, we have the functional equation
\begin{multline} \label{eqn:G-H-additive}
\GMat{(\alpha_{12}+ \alpha_{12}')\beta_{12}}{(\alpha_{12}+ \alpha_{12}')\beta_{21}}{(\alpha_{21} + \alpha_{21}')\beta_{12}}{(\alpha_{21} + \alpha_{21}')\beta_{21}} = \\
								  H(\alpha_{12}, \alpha_{21}) \, H(\beta_{12}, \beta_{21}) + H(\alpha'_{12}, \alpha'_{21}) \, H(\beta_{12}, \beta_{21}),
\end{multline}
valid for all~$(\alpha_{12}, \alpha_{21}), (\alpha'_{12}, \alpha'_{21}), (\beta_{12}, \beta_{21}) \in S$ for which the right-hand side lies in~$\cunitdisc$.

\begin{figure}
\begin{center}
\includegraphics[width=3in]{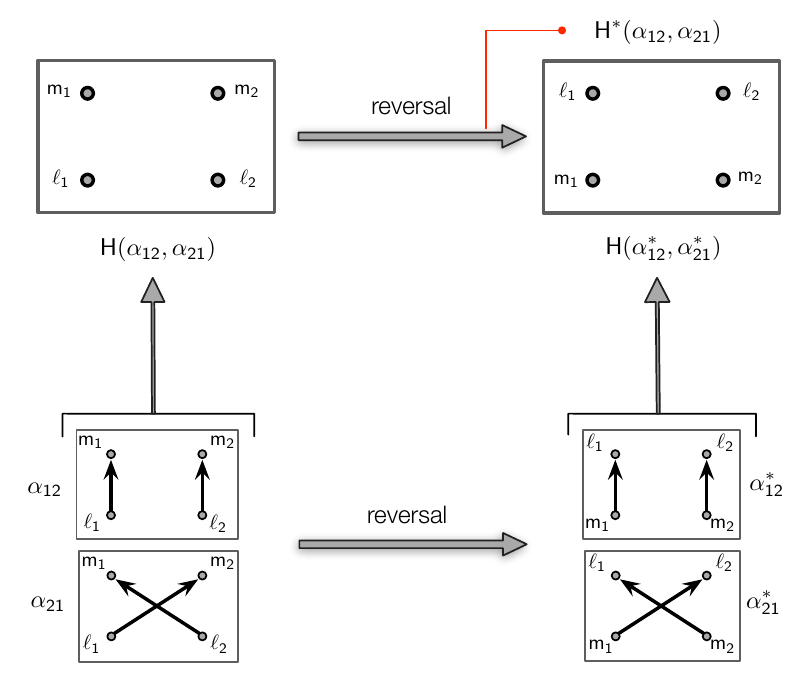}
\caption{\label{fig:reversal}  \emph{Reversal argument.}  On the top-right, a system of two indistinguishable particles are subject to measurement~$M$ first, and then~$L$ immediately afterwards, yielding outcomes~$m_1, m_2$ at time~$t_1$ and~$\ell_1, \ell_2$ at~$t_2$.  The amplitude of this process can be computed in two different ways.  As shown on the left, one can first take the amplitude,~$H(\alpha_{12}, \alpha_{21})$, of the indistinguishable-particle process with the measurements in their original order, and then take its complex-conjugate to get~$H^*(\alpha_{12}, \alpha_{21})$.  Alternatively, as shown on the bottom, one can compute the amplitude of the reversed process directly in terms of the distinguishable-particle sequences compatible with it.  This yields~$H(\alpha_{12}^*, \alpha_{21}^*)$.  Hence,~$H^*(\alpha_{12}, \alpha_{21}) = H(\alpha_{12}^*, \alpha_{21}^*)$.}
\end{center}
\end{figure}

Lastly, we consider an experiment on two identical particles in which the sequence of measurements performed on the particles is reversed, and the measurements follow one another immediately in time~(see Fig.~\ref{fig:reversal}).  %
As indicated in the figure, using the amplitude conjugation rule, the amplitude can be computed in two different ways, which can be equated to give
\begin{equation} \label{eqn:H-conjugation}
H^*(\alpha_{12}, \alpha_{21})  = H(\alpha_{12}^*, \alpha_{21}^*),
\end{equation}
valid for all~$(\alpha_{12}, \alpha_{21}) \in S$.

\section*{Solution of the functional equations}

We now show that the functional equations~\eqref{eqn:G-product}, \eqref{eqn:G-H-additive} and~\eqref{eqn:H-conjugation}, together with Eqs.~\eqref{eqn:normalization1} and~\eqref{eqn:normalization2}, lead to the symmetrization postulate for the two particle case.

Using Eq.~\eqref{eqn:G-product}, the left-hand side of Eq.~\eqref{eqn:G-H-additive} can be rewritten to give
\begin{multline*} \label{eqn:coarse-grain-amp2b}
H(\alpha_{12} + \alpha_{12}', \alpha_{21} + \alpha_{21}') \,H(\beta_{12}, \beta_{21})   =  \\ H(\alpha_{12}, \alpha_{21}) \, H(\beta_{12}, \beta_{21})  + H(\alpha'_{12}, \alpha'_{21}) \, H(\beta_{12}, \beta_{21}).
\end{multline*}
A nontrivial solution requires that~$H(\beta_{12}, \beta_{21})$ cannot be zero for all~$\beta_{12}, \beta_{21}$.  Hence, we obtain the additivity relation
\begin{equation} \label{eqn:H-additivity}
H(\alpha_{12} + \alpha_{12}', \alpha_{21} + \alpha_{21}')  =  H(\alpha_{12}, \alpha_{21}) +  H(\alpha_{12}', \alpha_{21}'),
\end{equation}
valid for all~$(\alpha_{12}, \alpha_{21}), (\alpha'_{12}, \alpha'_{21}), (\alpha_{12} + \alpha'_{12}, \alpha_{21} + \alpha'_{21}) \in S$.

We are now in a position to determine~$H$.  First, Eq.~\eqref{eqn:H-additivity} implies that
\begin{equation} \label{eqn:H-2-sum-A}
H(\alpha_{12}, \alpha_{21}')  =  H(\alpha_{12} + 0, 0 + \alpha_{21}') =  H(\alpha_{12}, 0) + H(0, \alpha_{21}').
\end{equation}
Now, Eqs.~\eqref{eqn:normalization1}  and~\eqref{eqn:normalization2} establish that~$(z,0), (0,z) \in S$ whenever~$z \in \cunitdisc$, and Eq.~\eqref{eqn:G-product} then implies that, for all~$z \in \cunitdisc$,
\begin{equation} \label{eqn:G-H-H}
\Gmat{0}{z}{0}{0} = H(z, 0) \, H(0, 1) = H(1, 0) \, H(0, z).
\end{equation}
Since~$H(1, 0) \neq 0$ from Eq.~\eqref{eqn:normalization1}, we can use Eq.~\eqref{eqn:G-H-H} to rewrite Eq.~\eqref{eqn:H-2-sum-A} as
\begin{equation} \label{eqn:H-2-sum-B}
H(\alpha_{12}, \alpha_{21}) = H(\alpha_{12}, 0) + \frac{H(0,1)}{H(1,0)} H(\alpha_{21}, 0).
\end{equation}

To determine the form of~$H(z, 0)$, we note from Eq.~\eqref{eqn:H-additivity} that, for all~$z_1, z_2, z_1 + z_2 \in \cunitdisc$,
\begin{equation} \label{eqn:two-H-add}
H(z_1 + z_2, 0) = H(z_1, 0) + H(z_2, 0),
\end{equation}
while Eq.~\eqref{eqn:G-product} implies that, for all~$z_1, z_2 \in \cunitdisc$,
\begin{equation} \label{eqn:two-H-multiply}
H(z_1 z_2, 0) \, H(1, 0) = H(z_1, 0) \, H(z_2, 0).
\end{equation}
Writing~$f(z) = H(z, 0)/H(1,0)$, Eqs.~\eqref{eqn:two-H-add} and~\eqref{eqn:two-H-multiply} can be written as a pair of functional equations,
\begin{subequations} \label{eqn:pair-functional-eqs}
\begin{align}
f(z_1 + z_2) 			&= 	f(z_1) + f(z_2) \label{eqn:pair-functional-eqs1}\\
f(z_1 z_2) 				&=	f(z_1) \, f(z_2), \label{eqn:pair-functional-eqs2}
\end{align}
\end{subequations}
where~$f(\cdot)$ is a complex-valued function of a complex argument.   We are interested in the continuous solutions of these equations for all~$z_1, z_2 \in \cunitdisc$ for which~$z_1 + z_2 \in \cunitdisc$.  As shown in Appendix~\ref{sec:functional}, these solutions are~$f(z) = z$, $f(z) = z^*$ or~$f(z) = 0$.  The zero solution is inadmissible since, from its definition,~$f(1) = 1$.  Therefore, 
\begin{equation*}
H(z, 0)= \begin{cases}
H(1,0) \,z  \\
H(1,0) \,z^*.
\end{cases}
\end{equation*}

From Eqs.~\eqref{eqn:normalization1} and~\eqref{eqn:normalization2}, we see that~$|H(1, 0)| = |H(0, 1)|$; and we note that Eq.~\eqref{eqn:H-conjugation} implies that~$H(1, 0)$ and~$H(0, 1)$ are both real.  Therefore, 
\begin{equation*}
H(0, 1) =  \pm  H(1, 0).
\end{equation*}
Hence, Eq.~\eqref{eqn:H-2-sum-B} becomes
\begin{subnumcases}{H(\alpha_{12}, \alpha_{21}) =} \label{eqn:two-particles-H-final2}
  H(1, 0) \left(  \alpha_{12} \pm  \alpha_{21} \right) \\
  H(1, 0) \left(  \alpha_{12} \pm  \alpha_{21} \right)^*. 
\end{subnumcases}
Since only the transition probability is of physical importance, only the modulus of~$H(\alpha_{12}, \alpha_{21})$ is relevant.  Hence, without loss of generality, we can take~$H(\alpha_{12}, \alpha_{21}) =  c\left(  \alpha_{12} \pm  \alpha_{21} \right)$, with~$c  = |H(1, 0)|$.   Finally, Eq.~\eqref{eqn:normalization1} fixes~$c=1$, leaving us with
\begin{equation} \label{eqn:two-particles-H-final}
H(\alpha_{12}, \alpha_{21}) =  \alpha_{12} \pm  \alpha_{21} ,
\end{equation}
where the sign is the only remaining degree of freedom, corresponding to bosons and fermions.  This completes the derivation of the symmetrization postulate for two identical, indistinguishable particles as expressed~\cite{FeynmanHibbs65} in Feynman's formulation of quantum theory.

The derivation for~$\nsys$ particles is given in Appendix~\ref{sec:many-identical}, and follows a similar line of argument, yielding the result
\begin{equation}\label{eqn:H-final}
H(\alpha_{\Bpi_1}, \alpha_{\Bpi_2}, \dots, \alpha_{\Bpi_{\nsys!}}) = \sum_{\Bpi \in \group{S}_\nsys} \left(\sgn(\Bpi)\right)^\sigma \,  \alpha_{\Bpi},
\end{equation}
where~$\sigma = 0, 1$ is the only remaining degree of freedom, whose value does not depend on the number of particles,~$N$.

\section*{Conclusion}

We have shown that there exists a direct path from the assumption that identical particles are indistinguishable to the symmetrization postulate.  Unlike the topological approach, the result applies to any quantum system, such as system of identical particles with spin, or a system of identical abstract finite dimensional subsystems, and relies on the standard quantum formalism rather than any variant thereof.  The derivation is a natural extension and application of the ideas previously used to derive Feynman's rules of quantum theory.  This demonstrates that the symmetrization postulate is comparable in nature and reliability to the core quantum formalism, contrary to what has previously appeared to be the case.  From the point of view of the derivation, the symmetrization postulate admits no natural variants, and, \emph{contra} the topological approach, is unrelated to the dimension of space.  In particular, the derivation implies that identical particles do not generically exhibit anyonic behavior in two dimensions.

As detailed elsewhere~\cite{Goyal-Knuth2011}, the methodology used to derive Feynman's rules has been previously used to derive~\cite{Cox-PT-paper, KnuthSkilling2012} the formalism of probability theory starting from Boolean logic, and has recently been applied to yield important insights in other domains~\cite{KnuthSkilling2012}.  The successful application of this methodology to the symmetrization postulate leads us to anticipate that this methodology may provide a valuable tool in understanding and constructing physical theory.

\begin{acknowledgments}
I thank Jonathan Burton, Jeremy Butterfield, Adam Caulton, Giulio Chiribella, Yiton Fu, Michael Hall, Klil Neori, Dean Rickles, Simon Saunders, John Skilling, and William Wootters for insightful discussions and comments.  This publication was, in part, made possible through the support of grants from the John Templeton Foundation and FQXi.
\end{acknowledgments}

\clearpage
\begin{widetext}
\begin{appendix}

\section{System of Many Identical Subsystems}
\label{sec:many-identical}

In this section, we present the derivation of the symmetrization postulate for~$N$ indistinguishable subsystems.  This derivation closely follows the derivation for two indistinguishable subsystems given in the main text.  For reasons of precision and clarity, we explicitly employ the the operational framework and notation used in Ref.~\cite{GKS-PRA}.  For completeness, we first summarize the key features of this framework, and then describe the extensions we need in order to describe composite systems.  We then summarize how one can describe Feynman paths and Feynman's amplitude rules in this operational framework.  Having laid this foundation, we present the derivation itself.

\subsection{Operational Framework}
\label{sec:framework}

As explained in Refs.~\cite{Goyal-QT2c, GKS-PRA}, the measurements and interactions which can be employed in a given experiment must be carefully circumscribed if they are to lead to a well-defined theoretical model.  The overarching requirement is that, in the given experiment, the outcome probabilities of any measurement but the first are independent of any interactions with the system prior to the first measurement, a requirement we refer to as \emph{experimental closure}.  Intuitively, the first measurement `screens off' the prior history of the system, rendering this prior history irrelevant insofar as making predictions about the outcomes of subsequent measurements in the experiment is concerned.  The requirement of experimental closure naturally leads to the following definitions and constraints.  

An experimental set-up is defined by specifying a source of physical systems, a sequence of measurements to be performed on a system on a run of the experiment, and the interactions with the system which occur between the measurements.   In a run of an experiment, a physical system from the source passes through a sequence of measurements~$\mment{L}, \mment{M}, \mment{N}, \dots$, which respectively yield outcomes~$\ell, m, n, \dots$ at times~$t_1, t_2, t_3, \dots$.  These outcomes are summarized in the \emph{measurement sequence} $\llseq{\ell}{m}{n}{\dots}$.  To avoid notational clutter, the measurements that yield these outcomes, the times at which these outcomes occur, and the nature of the system that is under examination, are all left implicit in the notation, and so must be inferred from the context.    In between these measurements, the system may undergo interactions with the environment.     

Over many runs of the experiment, the experimenter will observe the frequencies of the various possible measurement sequences, from which one can~(using Bayes' rule) estimate the probability associated with each sequence.   We define the probability~$\prob{A}$ associated with sequence~$\sseq{A} = \llseq{\ell}{m}{n}{\dots}$ as the probability of obtaining outcomes~$m, n, \dots$ conditional upon obtaining~$\ell$,
\begin{equation} \label{eqn:def-of-probability-of-sequence}
\prob{A} = \Pr(m, n, \dots \,|\, \ell).
\end{equation}

A particular outcome of a measurement is either \emph{atomic} or \emph{coarse-grained}.   An atomic outcome is one that cannot be more finely divided in the sense that the detector whose output corresponds to the outcome cannot be sub-divided into smaller detectors whose outputs correspond to two or more outcomes.  An example of atomic outcomes are the two possible outcomes of a Stern-Gerlach measurement performed on a silver atom.  A coarse-grained outcome is one that does not differentiate between two or more outcomes, an example being a Stern-Gerlach measurement where a detector's field of sensitivity encompasses the fields of sensitivity of two detectors each of which corresponds to a different atomic outcome.    Abstractly, if measurement~$\cmment{M}$ has an outcome which is a coarse-graining of the outcomes labeled~$1$ and~$2$ of measurement~$\mment{M}$, the outcome of~$\cmment{M}$ is labeled~$\bubble{1}{2}$, and this notational convention naturally extends to coarse-graining of more than two atomic outcomes.    In general, if all of the possible outcomes of a measurement are atomic, we shall call the measurement itself atomic.  Otherwise, we say it is a coarse-grained measurement.

It is important that all of the measurements that are employed in an experimental set-up come from the same \emph{measurement set},~$\set{M}$, or are coarsened versions of measurements in this set.  This ensures that all the measurements are probing the same aspect of the system.    The set is operationally defined as follows.  Measurement~$\mment{M}$ forms a \emph{measurement pair} with~$\mment{L}$ if in both (a) experiment~1, where measurement~$\mment{M}$ is performed on a given system immediately after it has been prepared using~$\mment{L}$, and~(b) experiment~2, where measurement~$\mment{L}$ is used to prepare the given system, and measurement~$\mment{M}$ is performed immediately afterwards, we have \emph{experimental closure}, namely that the outcome probabilities of the final measurement in both instances are independent of interactions with the system prior to the preparation.  The measurement set,~$\set{M}$, containing~$\mment{M}$ is the set of all measurements that form a measurement pair with~$\mment{M}$.  Interactions that occur in the period of time \emph{between} measurements are likewise selected from a set,~$\set{I},$ of possible interactions, which are such that they preserve closure when performed between any pair of measurements from~$\set{M}$.   Experimental closure requires that the first measurement,~$\mment{M}$, in an experiment is atomic; the remaining measurements must either lie in the same measurement set as~$\mment{M}$ or be coarsened versions of measurements in this set.   

\subsubsection{Composite Systems}

The operational framework sketched above is concerned with measurements that probe the entire system.  But, in the case of a composite system, it is possible to perform measurements that only probe particular subsystems of the system. For example, consider a system subject to position measurements at successive times.  If, at each of these times, two clicks are registered, we say that the system is a composite system of two particles.

More formally, consider an experimental arrangement in which, at each time, one measurement is  performed on each of the two subsystems of a composite system.   If, at time~$t_1$, measurements~$\mment{L_1}$ are performed on one subsystem and~$\mment{L_2}$ on the other, and simultaneously yield outcomes~$\ell_1$ and~$\ell_2$, respectively, we shall notate the outcome of the two measurements as~$\cout{\ell_1}{\ell_2}$.  If the measurement~$L_1$ and~$L_2$ are performed at time~$t_1$, and measurements~$M_1$ and~$M_2$ are subsequently performed at~$t_2$, the resulting measurement sequence can accordingly be written as~$\sseq{C} = \seq{\cout{\ell_1}{\ell_2}}{\cout{m_1}{m_2}}$.   To ensure closure, the measurements~$\mment{L_1}, \mment{L_2}, \mment{M_1}$, and~$\mment{M_2}$ must all lie in the same measurement set.

If it is possible to carry out measurements immediately before~$t_1$ and after~$t_2$ whose outcomes together determine whether the subsystem detected at~$\mment{M}_1$ originated at~$\mment{L}_1$ or at~$\mment{L}_2$, we say that these subsystems are distinguishable.  Otherwise we say they are indistinguishable, in which case the statement that subsystem~$1$ originated at~$\mment{L}_1$ and transitioned to, say, $\mment{M}_1$, is not, in general, operationally meaningful.

\subsection{Operationalization of Feynman's paths and Feynman's rules}

\label{sec:operational-logic}

Consider an experimental set-up in which a physical system is subject to successive measurements~$\mment{L}, \mment{M}, \mment{N}$ at successive times~$t_1, t_2, t_3$, with there possibly being interactions with the system in the intervals between those measurements.  Here and subsequently, we assume that the measurements and interactions in any such set-up are selected according to the constraints described above.  We summarize the outcomes obtained in a given run of the experiment as the \emph{sequence}~$\sseq{C} = \lseq{\ell}{m}{n}$.  This is the operational counterpart to a Feynman `path'.

We now wish to formalize the idea that set-ups are interrelated in particular ways.  In Ref.~\cite{GKS-PRA}, we considered three such relationships.  First, the above set-up could be viewed as a \emph{series concatenation} of two experiments, the first in which measurements~$\mment{L}$ and~$\mment{M}$ occur at times~$t_1$ and~$t_2$, yielding the sequence~$\sseq{A} = \seq{\ell}{m}$, and the second in which measurements~$\mment{M}$ and~$\mment{N}$ occur at times~$t_2$ and~$t_3$, yielding~$\sseq{B} = \seq{m}{n}$.  In order to ensure that experimental closure is satisfied in the second experiment, measurement~$\mment{M}$ must be atomic.  Formally, we express this concatenation as
\begin{equation}
\sseq{C} = \sseq{A} \ser \sseq{B},
\end{equation}
where~$\ser$ is the \emph{series} combination operator.  More generally, the series operator can be used to concatenate two sequences provided their initial and final measurements are atomic, and the final measurement and outcome of the first sequence is the same as the initial measurement and outcome of the second sequence.   

Second, one can consider a set-up which is identical to the one above, except that outcomes~$m$ and~$m'$ of~$\mment{M}$ have been coarse-grained, so that one obtains the sequence~$\sseq{E} = \lseq{\ell}{\bubble{m}{m'}}{n}$.  Formally, we express the relationship of this sequence to the sequences~$\sseq{C} = \lseq{\ell}{m}{n}$ and~$\sseq{D} = \lseq{\ell}{m'}{n}$ as
\begin{equation}
\sseq{E} = \sseq{C} \pll \sseq{D},
\end{equation}
where~$\pll$ is the \emph{parallel} combination operation.  More generally, the parallel operator can combine any two sequences which are identical except for differing in the outcome of a single measurement in the set-up, provided that this measurement is not the initial or final measurement.

Third and finally, we can consider a set-up which is identical to the one above, except that the sequence of measurements has been reversed, so that measurements~$\mment{N}, \mment{M}$ and~$\mment{L}$ occur at times~$t_1, t_2$ and~$t_3$, respectively, and yield the sequence~$\sseq{F} = \lseq{n}{m}{\ell}$.  We express the relationship of this sequence to~$\sseq{C}$ as
\begin{equation}
\sseq{F} = \revsseq{C}
\end{equation}
where~$\revsseq{\cdot}$ is the \emph{reversal} operator.

\subsubsection{Feynman's Rules in Operational Form}
\label{sec:Feynmans-rules}

From the above definitions, it follows that the operators~$\ser$ and~$\pll$ satisfy several symmetry relations, to which we collectively refer to as an \emph{operational logic}. In Ref.~\cite{GKS-PRA}, it is shown that Feynman's rules are the unique pair-valued representation of this logic consistent with a few additional assumptions.  Writing~$\amp({X})$ for the complex-valued \emph{amplitude} that represents sequence~$\sseq{X}$, one finds
\begin{align}
\amp(\sseq{A} \pll \sseq{B}) 	&= \amp(\sseq{A}) + \amp(\sseq{B}) 	\tag{amplitude sum rule}\\
\amp(\sseq{A} \ser \sseq{B}) 	&= \amp(\sseq{A}) \cdot \amp(\sseq{B}) \tag{amplitude product rule}\\
\amp(\revsseq{A}) 				&= \amp^*(\sseq{A})  \label{eqn:amp-conjugation} \tag{amplitude conjugation rule} \\
P(\sseq{A}) 		&= \left| \amp(\sseq{A}) \right|^2.		\tag{probability rule}
\end{align}
These are Feynman's rules for measurements on single quantum systems.  
We remark that the amplitude conjugation rule only holds in the special case where the measurements in sequence~$\sseq{A}$ immediately follow one another in time.

\subsection{Derivation}
\label{sec:derivation}

The derivation for~$N$ indistinguishable subsystems closely follows the derivation for two indistinguishable subsystems given in the main text.  The key postulates are as follows.  Consider an experiment involving~$N > 1$ indistinguishable subsystems, with measurements~$L_1, L_2, \dots, L_\nsys$ performed at time~$t_1$, and~$M_1, M_2, \dots, M_\nsys$ performed at~$t_2$.   We postulate that the amplitude of the sequence~$\sseq{A} = \seq{\ell}{m}$, where~$\ell \equiv \coutd{\ell_1}{\ell_2}{\ell_\nsys}$ and~$m \equiv \coutd{m_1}{m_2}{m_\nsys}$, is a function of the amplitudes of the transitions of~$N$ distinguishable particles compatible with~$\sseq{A}$.   There are~$N!$ such transitions, each corresponding to some permutation,~$\Bpi$, in the symmetric group~$\group{S}_\nsys$:~in a transition corresponding to~$\Bpi$, the subsystem that was measured at~$\mment{L}_j$ is later found at~$\mment{M}_{\Bpi(j)}$, for~$j = 1, 2, \dots, N$.  We notate this transition as~$\Bpi_1 \rightarrow \Bpi$, where~$\Bpi_1$ is the identity permutation here and subsequently.  Accordingly, we write
\begin{equation} \label{eqn:H-general}
\amp\left(\sseq{A}\right) = H(\alpha_{\Bpi_1}, \alpha_{\Bpi_2}, \dots, \alpha_{\Bpi_{\nsys!}}),
\end{equation}
where the~$\Bpi_i$ are the~$\nsys !$ distinct permutations in~$\group{S}_\nsys$, and~$\alpha_{\Bpi}$ is the amplitude of the distinguishable-particle transition~$\Bpi_1 \rightarrow \Bpi$.

Similarly, for three stages of measurement, with each stage consisting of~$\nsys$ measurements, we postulate that the sequence~$\sseq{C} = \lseq{\ell}{m}{n}$, where~$n \equiv \coutd{n_1}{n_2}{n_\nsys}$, has amplitude
\begin{equation} \label{eqn:G-general}
\amp\left(\sseq{C}\right) = G(\mat{\Gamma}),
\end{equation}
where~$\mat{\Gamma}_{ij} = \gamma_{ij}$ is the amplitude of the transition~$\Bpi_1 \rightarrow \Bpi_i \rightarrow \Bpi_j$.

\subsubsection{Isolation Condition.}
\label{sec:isolation-condition} 

Consider an experiment involving~$N$ indistinguishable subsystems, with measurements~$L_1, L_2, \dots, L_\nsys$ performed at time~$t_1$, and~$M_1, M_2, \dots, M_\nsys$ performed at~$t_2$.   If a subset of these subsystems is bound to an isolated subexperiment, we shall assume, as is conventional, that we can compute the transition probability for the subexperiment without regard for the subsystems that are not part of the subexperiment.  This is the isolation condition.

We shall consider two special cases.   In the first, each subsystem is localized to its own subexperiment~(see Fig.~\ref{fig:isolation-two-particles}).  As an example, we may have~$N$ electrons, each localized to a widely separated hydrogen atom.   In particular, suppose that, for~$j = 1, \dots, \nsys$, and some permutation~$\Bpi_i$, the subexperiment involving measurements~$L_j$ and~$M_{\Bpi_i(j)}$ is isolated.  That is, with certainty, the subsystem originating at outcome~$\ell_j$ of measurement~$L_j$ will later be later detected at some outcome,~$m_{\Bpi_i(j)}$, of measurement~$M_{\Bpi_i(j)}$, and the probability of the transition from~$\ell_j$ to~$m_{\Bpi_i(j)}$ is independent of the transition probabilities of the~$\nsys - 1$ other subexperiments.     

By the isolation condition,  we can compute the transition probability of the transition from~$\ell_j$ to~$m_{\Bpi_i(j)}$ without regard to the other subexperiments.  If the transition~$\ell_j$ to~$m_{\Bpi_i(j)}$  has amplitude~$u_j$ for~$j = 1,\dots, N$, then the corresponding transition probability is~$|u_j|^2$.  Hence, the probability of the transition from~$\ell$ to~$m$ is
\begin{equation} \label{eqn:isolation-case1-1}
\Pr(m\mid\ell) = \prod_{j=1}^N  |u_j|^2 = |z|^2,
\end{equation}
where~$z = u_1 u_2 \dots u_N$.   We can also compute this transition probability using Eq.~\eqref{eqn:H-general} with amplitude~$\alpha_{\Bpi_k} = z\,\delta_{\Bpi_i, \Bpi_k}$.  Now, let us define~$J_i(z) \equiv H(0, \dots, 0, z, 0, \dots, 0)$ where~$z$ appears as the~$i$th argument of~$H$.  Then, the amplitude of the transition is~$J_i(z)$, so that
\begin{equation} \label{eqn:isolation-case1-2}
\Pr(m\mid\ell) = |J_i(z)|^2.
\end{equation}
Consistency of Eqs.~\eqref{eqn:isolation-case1-1} and~\eqref{eqn:isolation-case1-2} requires that, for all~$i$,
\begin{equation} \label{eqn:modulus-H-of-P}
|J_i(z)| = |z|.
\end{equation}
%


In the second special case, we have two subsystems localized to their own subexperiment, while each of the other subsystems are localized to their own subexperiment~(see Fig.~\ref{fig:isolation-three-particles}).  As an example, we may have two electrons bound to a helium atom, while the other~$N-2$ electrons are each bound to their own hydrogen atom.
\begin{figure}
\begin{center}
\includegraphics[width=5.5in]{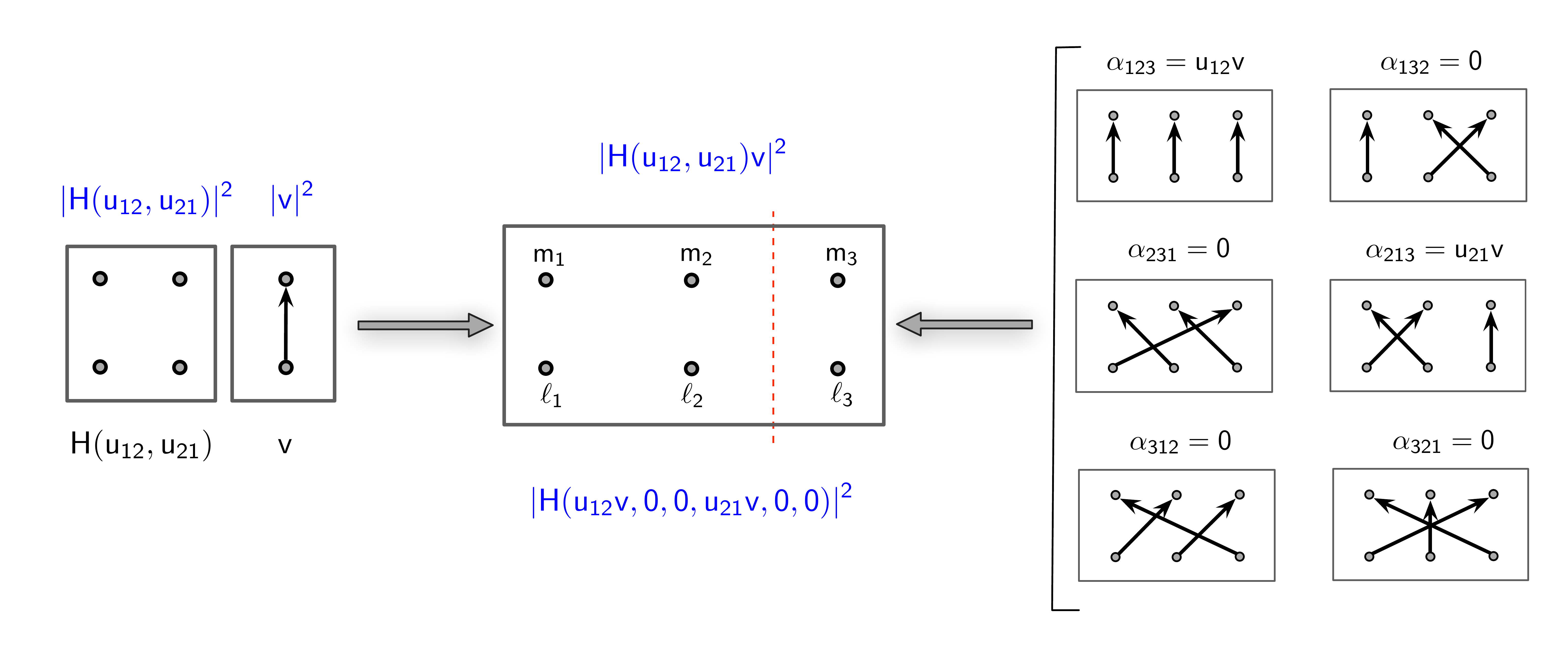}
\caption{\label{fig:isolation-three-particles}  \emph{Isolation condition (second special case).}  In the center, three indistinguishable particles are each subject to measurements at times~$t_1$ and $t_2$, yielding the indicated outcomes.  There are two subexperiments, one involving measurements~$L_1, L_2$ and~$M_1, M_2$, and the other involving measurements~$L_3$ and~$M_3$.   Accordingly, the transition probability can be computed in two different ways.  As on the left, one can compute the transition probabilities,~$|H(u_{12}, u_{21})|^2$ and~$|v|^2$, for each subexperiment, and multiply these to obtain~$|H(u_{12}, u_{21})v|^2$.  Alternatively, as on the right, one can use the~$H$ function to compute the amplitude~$H(\alpha_{123}, \alpha_{132}, \alpha_{231}, \alpha_{213}, \alpha_{312}, \alpha_{321}) = H(u_{12}v, 0, 0, u_{21}v, 0,	 0)$, which yields transition probability~$|H(u_{12}v, 0, 0, u_{21}v, 0 0)|^2$.  Consistency requires that~$|H(u_{12}v, 0, 0, u_{21}v, 0, 0)|^2 = |H(u_{12}, u_{21})v|^2$.  Setting~$v = 1$, and using the previously derived form~$H(u_{12}, u_{21}) = u_{12} \pm u_{21}$, we obtain~$|H(u_{12}, 0, 0, u_{21}, 0 0)| = |u_{12} \pm u_{21}|$.}
\end{center}
\end{figure}
Let us suppose that the measurements~$L_a, L_b$ and~$M_{\Bpi(a)}, M_{\Bpi(b)}$ form a subexperiment, while~$L_j$ and~$M_{\Bpi(j)}$ for~$j \neq a, b$, form a subexperiment, where~$\Bpi$ is some permutation.

By the isolation condition, we can compute the transition probability from~$\ell$ to~$m$ in two different ways.  First, the transition amplitude for the subexperiment involving~$L_a, L_b$ is~$H(u_{12}, u_{21})$, where~$u_{12}, u_{21}$ are the amplitudes of the direct and indirect transitions of two distinguishable systems from~$L_a, L_b$ to~$M_{\Bpi(a)}, M_{\Bpi(b)}$.  Therefore, the transition probability for this subexperiment is~$|H(u_{12}, u_{21})|^2$.    And let us suppose that the amplitudes of the transitions of remaining~$N-2$ subexperiments are all unity, so that their transition probabilities are all unity.  Then, the overall transition probability from~$\ell$ to~$m$ is~$|H(u_{12}, u_{21})|^2$.

Second, we can use~$H$ for~$N$ particles~(Eq.~\eqref{eqn:H-general}) with~$\alpha_{\Bpi_i} = u_{12} \delta_{\Bpi, \Bpi_i} + u_{21} \delta_{\Bpi', \Bpi_i}$, where~$\Bpi' = \Btau\Bpi$ and~$\Btau$ is the transposition~$(\Bpi(a), \Bpi(b))$, to obtain the amplitude~$H(\alpha_{\Bpi_1}, \alpha_{\Bpi_2}, \dots, \alpha_{\Bpi_{N!}})$, and hence the transition probability~$|H(\alpha_{\Bpi_1}, \alpha_{\Bpi_2}, \dots, \alpha_{\Bpi_{N!}})|^2$ for the overall transition.  Consistency requires that
\begin{equation}
|H(\alpha_{\Bpi_1}, \alpha_{\Bpi_2}, \dots, \alpha_{\Bpi_{N!}})| 	= |H(u_{12}, u_{21})|.
\end{equation}
Using Eq.~\eqref{eqn:two-particles-H-final}, this becomes
\begin{equation} \label{eqn:isolation-general}
|H(\alpha_{\Bpi_1}, \alpha_{\Bpi_2}, \dots, \alpha_{\Bpi_{N!}})|  = |u_{12} + (-1)^\sigma u_{21}|
\end{equation}
where~$\sigma = 0$ or~$1$, with~$\alpha_{\Bpi_i} = u_{12} \delta_{\Bpi, \Bpi_i} + u_{21} \delta_{\Bpi', \Bpi_i}$.

\subsubsection{The $G$-product equation.}

In the case where the outcomes at~$t_2$ are atomic, we can compute the amplitude of the sequence~$\sseq{C} =\lseq{\ell}{m}{n}$ in two different ways.  First, writing~$\alpha_{\Bpi}$ as the amplitude of the distinguishable-particle transition~$\Bpi_1 \rightarrow \Bpi$ compatible with~$\seq{\ell}{m}$ and similarly~$\beta_{\Bpi}$ as the amplitude of the transition~$\Bpi_1 \rightarrow \Bpi$ compatible with~$\seq{m}{n}$,  we can directly use Eq.~\eqref{eqn:G-general} to get
\begin{equation*}
\amp(\sseq{C}) = G(\mat{\Gamma}),
\end{equation*}
where~$\mat{\Gamma}_{ij} = \alpha_{\Bpi_i} \, \beta_{\Bpi_j}$ is the amplitude of the distinguishable-particle transition~$\Bpi_1 \rightarrow \Bpi_i \rightarrow \Bpi_j$, computed using the amplitude product rule.  Alternatively, we can write
\begin{equation*}
\sseq{C} = \seq{\ell}{m} \,\ser\, \seq{m}{n}
\end{equation*}
so that
\begin{equation*}
\amp(\sseq{C}) = H(\alpha_{\Bpi_1}, \dots, \alpha_{\Bpi_{\nsys!}} ) \,H(\beta_{\Bpi_1}, \dots, \beta_{\Bpi_{\nsys!}} ).
\end{equation*}
Equating these two results, we obtain a functional equation, the $G$-product equation,
\begin{equation} \label{eqn:G-product-general}
G(\mat{\Gamma}) =  H(\alpha_{\Bpi_1}, \dots, \alpha_{\Bpi_{\nsys!}} ) \,H(\beta_{\Bpi_1}, \dots, \beta_{\Bpi_{\nsys!}} ).
\end{equation}

\subsubsection{Coarse graining, and the additivity relation for $H$.}

Next, consider an experiment with measurements at times~$t_1$,~$t_2$ and~$t_3$ where measurement~$M_q$, for some~$q \in \{1, \dots, \nsys\}$, is coarse-grained, and has an outcome is~$\tilde{m}_q = (m^{(1)}_q, m^{(2)}_q)$.  Suppose that the sequence~$\sseq{C} = \lseq{\ell}{m}{n}$, where~$m \equiv \coutdd{m_1}{m_2}{\tilde{m}_q}{m_{\nsys}}$,  is obtained.  We can compute its amplitude in two ways.  

Let~$\alpha^{(r)}_{\Bpi_i}$ be the amplitude of the transition~$\Bpi_1 \rightarrow \Bpi_i$ compatible with~$\seq{\ell}{\coutdd{m_1}{m_2}{m_q^{(r)}}{m_\nsys}}$, and likewise~$\beta^{(r)}_{\Bpi_j}$ the amplitude of the transition~$\Bpi_1 \rightarrow \Bpi_j$ compatible with~$\seq{\coutdd{m_1}{m_2}{m_q^{(r)}}{m_\nsys}}{n}$.  We restrict attention to the special case where~$\beta^{(r)}_{\Bpi_j} = \beta_{\Bpi_j}$.    Then, we can directly apply Eq.~\eqref{eqn:G-general} to get
\begin{equation*}
\amp({\sseq{C}}) = G(\mat{\Gamma}),
\end{equation*}
where~$\Gamma_{ij} = (\alpha^{(1)}_{\Bpi_i} + \alpha^{(2)}_{\Bpi_i})\beta_{\Bpi_j}$ is computed using the amplitude sum and product rules.  Using the $G$-product equation~(Eq.~\eqref{eqn:G-product-general}), this can be written
\begin{equation*}
\amp({\sseq{C}}) = H(\alpha^{(1)}_{\Bpi_1} + \alpha^{(2)}_{\Bpi_1}, \dots, \alpha^{(1)}_{\Bpi_{\nsys!}} + \alpha^{(2)}_{\Bpi_{\nsys!}}) \, H(\beta_{\Bpi_1}, \dots, \beta_{\Bpi_{\nsys!}}).
\end{equation*}

Alternatively, we can write the sequence~$\sseq{C}$ as
\begin{equation*}
\sseq{C} = \bigvee_{r=1}^2 \lseq{\ell}{\coutdd{m_1}{m_2}{m_q^{(r)}}{m_\nsys}}{n} 
\end{equation*}
whose amplitude is given using the amplitude sum rule by
\begin{equation*}
\amp(\sseq{C}) =
\sum_{r = 1}^{2} H	(\alpha^{(r)}_{\Bpi_1}, \alpha^{(r)}_{\Bpi_2}, \dots, \alpha^{(r)}_{\Bpi_{\nsys!}})
						\, H(\beta_{\Bpi_1}, \dots, \beta_{\Bpi_{\nsys!}}).
\end{equation*}

If~$H(\beta_{\Bpi_1}, \dots, \beta_{\Bpi_{\nsys!}}) \neq 0$ for some values of~$\beta_{\Bpi_1}, \dots, \beta_{\Bpi_{\nsys!}}$, we can equate these two expressions for~$\amp(\sseq{C})$ to obtain the additivity relation
\begin{equation} \label{eqn:H-additivity-general}
H(\alpha^{(1)}_{\Bpi_1} + \alpha^{(2)}_{\Bpi_1}, \alpha^{(1)}_{\Bpi_2} + \alpha^{(2)}_{\Bpi_2}, \dots, \alpha^{(1)}_{\Bpi_{\nsys!}} + \alpha^{(2)}_{\Bpi_{\nsys!}}) = \sum_{r = 1}^{2} H	(\alpha^{(r)}_{\Bpi_1}, \alpha^{(r)}_{\Bpi_2}, \dots, \alpha^{(r)}_{\Bpi_{\nsys!}}).
\end{equation}
The case~$H(\beta_{\Bpi_1}, \dots, \beta_{\Bpi_{\nsys!}}) = 0$ for all~$\beta_{\Bpi_1}, \dots, \beta_{\Bpi_{\nsys!}}$ is the trivial solution that implies that every transition of~$N$ indistinguishable particles has zero probability.  This solution can therefore be discarded.

\subsubsection{Simplification of expression for~$H$.}

From the additivity relation Eq.~\eqref{eqn:H-additivity-general}, it follows that
\begin{equation*}
H(\alpha_{\Bpi_1} + 0, \alpha_{\Bpi_2} + 0, \dots, \alpha_{\Bpi_{\nsys! - 1}} + 0, 0 + \alpha_{\Bpi_{\nsys!}}) = H(\alpha_{\Bpi_1}, \alpha_{\Bpi_2}, \dots, \alpha_{\Bpi_{\nsys! - 1}}, 0)
+ H(0, 0, \dots, 0, \alpha_{\Bpi_{\nsys!}}).
\end{equation*}
Iteratively applying the additivity relation, we obtain
\begin{equation} \label{eqn:H-sum}
\begin{aligned}
H(\alpha_{\Bpi_1}, \alpha_{\Bpi_2}, \dots, \alpha_{\Bpi_{\nsys!}}) 	&= H(\alpha_{\Bpi_1}, 0, \dots, 0) + H(0, \alpha_{\Bpi_2}, 0, \dots, 0) + \dots + H(0, 0, \dots, 0, \alpha_{\Bpi_{\nsys!}}) \\
																									&= J_{1}(\alpha_{\Bpi_1}) + J_{2}(\alpha_{\Bpi_2}) + \dots + J_{{\nsys!}}(\alpha_{\Bpi_{\nsys!}}).
\end{aligned}
\end{equation}

We can interrelate the~$J_{i}(\cdot)$ using the $G$-product equation (Eq.~\eqref{eqn:G-product-general}).  Taking~$\Gamma_{ij} = \delta_{ik} \delta_{j\ell} \,z$, and noting that Eq.~\eqref{eqn:modulus-H-of-P} ensures that all~$z \in \cunitdisc$ lie in the domain of~$J_i$ for all~$i$, the $G$-product equation implies
\begin{equation*}
G(\mat{\Gamma}) = J_{k}(z) \, J_{\ell} (1) = J_{k}(1) \, J_{\ell} (z).
\end{equation*}
Let us write~$J(z) \equiv J_{1}(z)$.  Since~$J(1) \neq 0$ from Eq.~\eqref{eqn:modulus-H-of-P}, we can use the above expression for~$G(\mat{\Gamma})$ to rewrite Eq.~\eqref{eqn:H-sum} solely in terms of~$J(z)$ and the constants~$J_{i}(1)$:
\begin{equation*} 
H(\alpha_{\Bpi_1}, \alpha_{\Bpi_2}, \dots, \alpha_{\Bpi_{\nsys!}}) 	=  \frac{1}{J(1)} \sum_{i=1}^{\nsys!}  J_i(1) J(\alpha_{\Bpi_i}).
\end{equation*}

To fix the form of~$J(z)$, we note that, from Eq.~\eqref{eqn:H-additivity-general},
\begin{equation} \label{eqn:J-add}
J(z_1 + z_2)  = J(z_1) + J(z_2),
\end{equation}
while Eq.~\eqref{eqn:G-product-general} implies
\begin{equation} \label{eqn:J-multiply}
J(z_1z_2) \, J(1) = J(z_1) \, J(z_2).
\end{equation}
Writing~$J(z) = J(1) \, f(z)$ transforms Eqs.~\eqref{eqn:J-multiply} and~\eqref{eqn:J-add} into Eqs.~\eqref{eqn:pair-functional-eqs1} and~\eqref{eqn:pair-functional-eqs2}.  As in the two-particle case in the main text, we are interested in the continuous solutions of these equations for all those~$z_1, z_2$ which correspond to probabilities in~$[0,1]$ for which~$z_1 + z_2, f(z_1), f(z_2)$ and~$f(z_1 + z_2)$ also correspond to probabilities in~$[0,1]$.  Now, due to Eq.~\eqref{eqn:modulus-H-of-P}, $J(z)$, and so also~$f(z)$, lies in the closed unit disc,~$\cunitdisc$, in the complex plane whenever~$z \in\cunitdisc$.  Therefore, we seek the continuous solutions of Eqs.~\eqref{eqn:pair-functional-eqs1} and~\eqref{eqn:pair-functional-eqs2} for all~$z_1, z_2 \in \cunitdisc$ for which~$z_1 + z_2 \in \cunitdisc$.  As shown in Appendix~\ref{sec:functional}, these solutions are~$f(z) = z$, $f(z) = z^*$ or~$f(z) = 0$.  The zero solution is inadmissible since, from its definition,~$f(1) = 1$.  Hence, 
\begin{subnumcases}{J(z)=}
J(1) \,z  \\
J(1) \,z^*.
\end{subnumcases}
Therefore, writing~$J_i(1)$ as~$Q(\Bpi_i)$,
\begin{subnumcases}{\label{eqn:N-particles-H-final1}H(\alpha_{\Bpi_1}, \alpha_{\Bpi_2}, \dots, \alpha_{\Bpi_{\nsys!}})  =} 
\sum_{\Bpi \in \group{S}_\nsys}  Q(\Bpi) \, \alpha_{\Bpi} \\
\sum_{\Bpi \in \group{S}_\nsys} Q(\Bpi) \, \alpha^*_{\Bpi}.
\end{subnumcases}

\paragraph{Determination of values of the~$Q(\Bpi)$.}

To determine the values of~$Q(\Bpi)$ for all possible permutations~$\Bpi \in \group{S}_n$, we first show that~$Q(\Bpi) = \pm 1$, and then show that~$Q(\Bpi) = Q(\Bpi')$ whenever~$\Bpi$ and~$\Bpi'$ are both odd or both even.

\subparagraph{i. Establishing that~$Q(\Bpi) =  \pm 1 $.}

The reverse,~$\revsseq{\sseq{A}}$, of sequence~$\sseq{A} = \seq{\ell}{m}$, where~$\ell \equiv \coutd{\ell_1}{\ell_2}{\ell_\nsys}$ and~$m \equiv \coutd{m_1}{m_2}{m_\nsys}$ can be computed in two distinct ways~(see Fig.~\ref{fig:reversal}) if the measurements immediately follow one another in time.  Equating the resulting expressions, we obtain
\begin{equation*}
H(\alpha^*_{\Bpi_1}, \alpha^*_{\Bpi_2}, \dots, \alpha^*_{\Bpi_{\nsys!}}) = H^*(\alpha_{\Bpi_1}, \alpha_{\Bpi_2}, \dots, \alpha_{\Bpi_{\nsys!}}).
\end{equation*}
In particular, for all~$\Bpi$,
\begin{equation*}
Q(\Bpi) = Q^*(\Bpi),
\end{equation*}
which, together with Eq.~\eqref{eqn:modulus-H-of-P}, implies
\begin{equation} \label{eqn:H-P-is-pm1}
Q(\Bpi) = \pm 1.
\end{equation}

\subparagraph{ii. Establishing~$Q(\Bpi) = Q(\Bpi')$ whenever~$\Bpi$ and~$\Bpi'$ are both odd or both even.}

In Eq.~\eqref{eqn:isolation-general}, let~$\alpha_{\Bpi_i} = k(\delta_{\Bpi, \Bpi_i} + \delta_{\Bpi', \Bpi_i})$, where~$k$ is some constant.  Then, using Eq.~\eqref{eqn:N-particles-H-final1}, Eq.~\eqref{eqn:isolation-general} becomes
\begin{equation*}
|Q(\Bpi)  + Q(\Bpi')| = |1 + (-1)^\sigma|,
\end{equation*}
with~$\sigma = 0$ or~$1$, where~$\Bpi$ is any permutation and~$\Bpi'$ is of the form~$\Btau\Bpi$ where~$\Btau$ is any transposition.  Therefore,
\begin{equation} \label{eqn:Q-bpi1}
Q(\Bpi') = (-1)^\sigma Q(\Bpi).
\end{equation}

Now, let~$\Bpi'' = \Btau' \Bpi'$, where~$\Btau'$ is some transposition.  Then, Eq.~\eqref{eqn:Q-bpi1} implies that~$Q(\Bpi'') = (-1)^\sigma Q(\Bpi')$.  Combining this with~$Q(\Bpi') = (-1)^\sigma Q(\Bpi)$, we obtain
\begin{equation}
Q(\Bpi'') = Q(\Bpi).
\end{equation}
That is, any two permutations,~$\Bpi, \Bpi''$, that are connected by a pair of transpositions have the same $Q$-value.
But every even permutation can be written as a product of an even number of transpositions and the identity permutation.  Therefore, all even permutations have the same $Q$-value,~$Q_e$.  Similarly, every odd permutation can be written as a product of an even number of transpositions and a given odd permutation.  Therefore, all odd permutations have the same $Q$-value,~$Q_o$.   Finally, Eq.~\eqref{eqn:Q-bpi1} implies that~$Q_o = (-1)^\sigma Q_e$.

With these results for~$Q(\Bpi)$, Eq.~\eqref{eqn:N-particles-H-final1} becomes
\begin{subnumcases}{\label{eqn:H-odd-even}H(\alpha_{\Bpi_1}, \alpha_{\Bpi_2}, \dots, \alpha_{\Bpi_{\nsys!}}) = \pm}
\sum_{\Bpi \in \group{S}_\nsys} \left(\sgn(\Bpi)\right)^\sigma  \alpha_{\Bpi} \\
\sum_{\Bpi \in \group{S}_\nsys} \left(\sgn(\Bpi)\right)^\sigma  \alpha^*_{\Bpi},
\end{subnumcases}
where~$\sgn(\Bpi)$ takes the value~$+1$ or~$-1$ according to whether~$\Bpi$ is even or odd.

Insofar as the probability of the transition of the system of~$N$ indistinguishable particles is concerned, the overall sign of~$H$ and the complex conjugation are irrelevant.   More generally, consider a system,~$\sys$, that consists of subsystems~$\sys_1$ and~$\sys_2$, where~$\sys_1$ consists of~$N$ indistinguishable particles of one type~(say, electrons).  Suppose, first, that~$\sys_2$ only contains particles that can be distinguished from those in~$\sys_1$.   As described above, let measurements~$\mment{L}_1, \dots, \mment{L}_N$ and~$\mment{M}_1, \dots, \mment{M}_N$ be performed on~$\sys_1$ at times~$t_1$ and~$t_2$.  Additionally, let measurements~$\mment{U}$  and~$\mment{V}$ be performed on~$\sys_2$ at~$t_1$ and~$t_2$, respectively, yielding outcomes~$u$ and~$v$.  Let~$\tilde{\alpha}_{\Bpi}$ be the amplitude of the transition of~$\sys$ from~$\cout{\ell}{u}$ to~$\cout{m}{v}$ in which the particles in~$\sys_1$ are treated as distinguishable and make the transition described by~$\Bpi$.  Then, by the same argument as described above, the amplitude of the process from~$\cout{\ell}{u}$ to~$\cout{m}{v}$ where the particles in~$\sys_1$ are treated as indistinguishable is given by~$H(\tilde{\alpha}_{\Bpi_1}, \tilde{\alpha}_{\Bpi_2}, \dots, \tilde{\alpha}_{\Bpi_{\nsys!}})$, and the transition probability is again unaffected by the overall sign or complex conjugation of~$H$.  In the case that~$\sys_2$ \emph{does} contain, say,~$M$ particles that are indistinguishable from those in~$\sys_1$, the boundaries of~$\sys_1$ must be redrawn to encompass them.  The resulting situation, namely a system composed of subsystem~$\sys_1'$, containing~$N+M$ indistinguishable particles, and subsystem~$\sys_2'$, containing only particles that are distinguishable from those in~$\sys_1'$, is of the same type the as one previously considered.  

Therefore, in general, the overall sign of~$H$ and the complex conjugation are irrelevant insofar as predictions are concerned, and can be discarded without any loss of generality.  Hence,
\begin{equation}\label{eqn:H-final}
H(\alpha_{\Bpi_1}, \alpha_{\Bpi_2}, \dots, \alpha_{\Bpi_{\nsys!}}) = \sum_{\Bpi \in \group{S}_\nsys} \left(\sgn(\Bpi)\right)^\sigma \,  \alpha_{\Bpi},
\end{equation}
where~$\sigma = 0$ or~$\sigma = 1$ is the only remaining degree of freedom, corresponding respectively to bosons and fermions.


\section{Solution of a pair of functional equations.}
\label{sec:functional}

We solve Eqs.~\eqref{eqn:pair-functional-eqs1} and~\eqref{eqn:pair-functional-eqs2} under the condition that~$z_1, z_2, z_1 + z_2 \in \cunitdisc$ with the aid of one of Cauchy's standard functional equations,
\begin{equation} \label{eqn:Cauchy}
h(x_1 + x_2) = h(x_1) + h(x_2),
\end{equation}
where~$h$ is a real function and~$x_1, x_2 \in [-a, a]$ with~$a \in \numberfield{R}$.  Its continuous solution is~$h(x) = \gamma x$ with~$\gamma \in\numberfield{R}$~\cite{Aczel-lectures-functional-equations}.

We first consider the solution of Eq.~\eqref{eqn:pair-functional-eqs1} under the condition that~$z_1, z_2$ lie in the closed half-unit circle,~$\chunitdisc$, which automatically ensures that~$z_1 + z_2$ lies in~$\cunitdisc$.  
Setting~$z_1 + z_2 = x + iy$, with~$x, y \in [-1/2, 1/2]$, in Eq.~\eqref{eqn:pair-functional-eqs1} gives
\begin{equation*} \label{eqn:f-decomp}
f(x + iy) = f(x) + f(iy).
\end{equation*}
Applying Eq.~\eqref{eqn:pair-functional-eqs1} again on~$f(x_1 + x_2)$ and~$f(iy_1 + iy_2)$, with~$x_1, x_2, y_1, y_2 \in [-1/2, 1/2]$ then implies 
\begin{align*}
f(x_1 + x_2) &= f(x_1) + f(x_2)  \\
f\left(i y_1 + i y_2 \right) &= f(iy_1) + f(iy_2).
\end{align*}
The real and imaginary parts of both of these equations each have the form of Eq.~\eqref{eqn:Cauchy}, and therefore have solutions
\begin{equation*}
f(x) = \alpha x  \quad\quad\text{and}\quad\quad f(iy) = \beta y
\end{equation*}
with~$\alpha, \beta \in \numberfield{C}$, so that, for all~$(x + iy) \in \chunitdisc$,
\begin{equation} \label{eqn:f-form-after-additivity}
f(x+iy) = \alpha x + \beta y.
\end{equation}
Since this equation satisfies Eq.~\eqref{eqn:pair-functional-eqs1} under the condition~$z_1, z_2, z_1 + z_2 \in \cunitdisc$, it is the general solution under these conditions.

From Eq.~\eqref{eqn:pair-functional-eqs2}, considered under the condition~$z_1, z_2, z_1 + z_2 \in \cunitdisc$,
\begin{equation*}
f(1 \cdot 1) = f(1) f(1)  \quad\quad\text{and}\quad\quad f(i \cdot i) = f(i) f(i),
\end{equation*}
which, due to Eq.~\eqref{eqn:f-form-after-additivity}, imply that
\begin{equation*}
\alpha = \alpha^2   \quad\quad\text{and}\quad\quad  {-\alpha} = \beta^2.
\end{equation*}
These have solutions~$(\alpha, \beta) = (0,0), (1,  i)$ and~$(1, -i)$, which correspond to~$f(z) = 0$, $f(z) = z$ and~$f(z) = z^*$.

\end{appendix}
\clearpage
\end{widetext}

\end{document}